\documentclass[proof]{WileyASNA-v1}

\articletype{Article Type}%

\received{17 January 2024}
\revised{24 January 2024}
\accepted{12 February 2024}

\begin{document}

\title{Spectroscopic and Photometric Behaviour of  LP~Ori}

\author[1,2]{A. Elmasl{\i}}

\author[1,2]{K. \"{O}. \"{U}nal} 

\author[1,2]{D. \"{O}zuyar}

\authormark{A. Elmasl{\i} \textsc{et al}}

\address[1]{\orgdiv{Ankara University}, \orgname{Science Faculty, Department of Astronomy and Space Sciences}, \orgaddress{\state{Ankara}, \country{T\"{u}rkiye}}}

\address[2]{\orgdiv{Ankara University}, \orgname{Astronomy and Space Sciences Research and Application Center (Kreiken Observatory), İncek Blvd., TR-06837, Ahlatlıbel}, \orgaddress{\state{Ankara}, \country{T\"{u}rkiye}}}

\corres{*Elmasl{\i}, Ankara University, Science Faculty, Department of Astronomy and Space Sciences, Ankara, T\"{u}rkiye. \email{elmasli@ankara.edu.tr}}

\presentaddress{Ankara University, Science Faculty, Department of Astronomy and Space Sciences, Ankara, T\"{u}rkiye}

\abstract{We performed spectroscopic and photometric analyses on the early B-type LP~Ori young stellar object located in the Orion Nebula. The high-resolution spectra of LP Ori was obtained at the Ankara University Kreiken Observatory in 2023, while all spectra recorded over the past 19 years were extracted from the ESO and ESPaDOnS archives. In these spectra of LP~Ori,  there is typically an emission observed in the core of the Balmer profile. This structure is accompanied by a $\sim$14-year interval inverse P-Cygni repetition superimposed on the Balmer profile. Additionally, an emission in the $\lambda$5875 He~I line is visible in the spectra in the year 2023. When this emission is considered together with the inverse P-Cygni structures, these observations suggest that LP Ori is a Herbig Ae/Be star. The abundance pattern of LP~Ori is close to solar with the exception of a slightly rich helium and slightly poor Al abundance. Additionally, the spectral energy distribution of LP~Ori was constructed to confirm the infrared excess caused by its circumstellar disk. Furthermore, the photometric analysis performed on the TESS observations of LP~Ori shows significant photometric variability and the frequency analysis reveals a $\beta$~Cephei star in its forthcoming evolution.}

\maketitle


\section{Introduction}\label{sec1}

LP~Ori is located in the Orion Nebula (M~42), which is a component of the larger Orion star-forming complex. \citet{odell01} identified a prominent bright reflection nebula around LP~Ori on the WFPC2 images captured by \textit{Hubble Space Telescope}. \citet{manoj02} emphasized that diffuse H\,II region is projected on to the line of sight to LP~Ori. Furthermore, the $^{12}$CO and $^{13}$CO radio observations of \citet{feddersen18} revealed that LP~Ori is centered in the expanding spherical named Shell\,11. 

 \citet{manoj02} remarked that little near-infrared excess in the spectral energy distribution (SED) is visible and  H$_\alpha$ emission was not detected in the spectra of LP~Ori. With these two facts \citet{manoj02} could not classify LP~Ori as a Herbig Ae/Be star and instead attributed it as one of the youngest Vega-like stars. \citet{fitzpatrick07} analysed the SED of LP~Ori across the infrared (IR) to ultraviolet (UV) wavelengths and determined a metallicity of $-0.14$~dex. \citet{petit08} discovered a magnetic field around LP~Ori and additional confirmation was obtained through spectropolarimetric observations performed by \citet{petit12} and \citet{2013MNRAS.429.1001A}. \citet{romanyuk13} asserted that LP~Ori is a helium-rich star with a magnetic field.  

In the photometric study conducted by \citet{2022MNRAS.515..998S}, the period of LP~Ori was determined to be 0.744 days (1.344 d$^{-1}$). This value was attributed to the rotation period of the star. \citet{2021A&A...654A..36S} also analysed the long cadence data obtained from the 6$^{\mathrm{th}}$ sector of the TESS satellite. As a result, they determined that the star had a significant frequency at 0.164 days (6.087 d$^{-1}$) with an amplitude of 0.41 milli-magnitudes and a SNR of 11.11. \citet{2018MNRAS.475.5144S} stated that the light curve has a period of 1.8551(5) days (0.539 d$^{-1}$) from frequency analysis. Besides, using the magnetic field changes, they found a period of 0.425 days (2.352 d$^{-1}$).

Herbig Ae/Be stars are pre-main sequence (PMS) stars with distinctive features seen in their visible and infrared spectrum. \citet{herbig60} investigated the visual spectra of 26 Be- and Ae-type stars and noticed the existence of  hydrogen Balmer emission lines, due to weak overlying shells. Later, these objects' evolutionary status were assigned as PMS with intermediate mass, leading to such objects classification as "Herbig Ae/Be stars".

The spectra of Herbig Ae/Be stars in the optic region exhibit emissions superimposed within the core of hydrogen Balmer profiles, along with emissions within specific (such as He\,I $\lambda$5875, 6678 and O\,I 7774\AA) absorption lines and also forbidden nitrogen and sulfur lines. All of these mentioned emission features in the spectra are also an outcome of the H\,II regions in star forming complexes. When  a star is situated above the surface of a molecular cloud, an ionization front materializes across its surface, resulting in the formation of a H\,II region that bears semblance to a blister, as in M\,42.

The Balmer emission line emanating from Herbig Ae/Be stars are generally broader and asymmetric profiles, in contrast to the  narrower and symmetric Balmer emission profiles observed in H\,II regions. The Herbig Ae/Be stars Hen\,3-1121S and Hen\,3-1121N \citet{carmona10}, are exceptions because they exhibit weak emission components with equivalent widths of 0.05 and 0.3 Å, respectively.

Herbig Ae/Be stars frequently undergo changes in the shape of their Balmer line profiles over time intervals. \citet{reipurth96} classified the characteristic features of Balmer emission lines into four types. Type-I corresponds to a symmetric emission-line profile, Type-II involves two peaks with the latter being stronger than the former, Type-III exhibits two peaks with the latter being weaker than the former, and Type-IV is characterized by the visibility of P-Cygni structures in the Balmer line profiles.

Herbig Ae/Be stars are known to have a strong IR-excess based on the presence of circumstellar matter. IR-excess profiles are divided into two groups according to the flat or ascending continuum \citep{2001A&A...365..476M}. This difference has a geometric origin due to the existence of inner and outer disk \citep{2005A&A...434..971D} as well as gaps in the disk \citep{2015ApJ...804..143H}. Indeed, near-IR scattering images of Herbig Ae/Be stars show large disk gaps that may be the basis for differences in SED in the IR region. Whether this is due to environmental factors or an evolutionary effect is still a matter of debate.

 Herbig Ae/Be stars show irregular photometric changes of an order of magnitude in the optical region, occurring to a characteristic time scale, generally from several days to weeks \citep{2002A&A...384.1038E, 2001A&A...379..564O, 1998ARA&A..36..233W}. It is also known that these changes are principally a result of photospheric activity and interplay of the star and the circumstellar environment. Especially rotating circumstellar disks, the rotational effect of cold photospheric spots or stellar pulsations are thought to be responsible from the variations \citep{1998ApJ...507L.141M}. 

As an extreme example, non-periodic photometric variations with amplitudes up to two-three magnitudes are observed in UX Ori-type stars. Many of these stars are catalogued as Herbig Ae/Be. Their extraordinary changes are explained by the obscuration of light from a source observed from the edge-on by a dusty cloud and scattering radiation in the circumstellar environment \citep{2018A&A...620A.128V}.

\section{Spectroscopic Observations and Reductions} 

LP~Ori was observed on the $25^{th}$ of February 2023 by selecting the high-resolution (R$\sim$30,000) mode on the Whoppshel\footnote{https://www.shelyak.com/produit/whoppshel/}
échelle spectrograph attached to the 0.8~m  Prof. Dr. Berahitdin Albayrak telescope (T80) at the Ankara University Kreiken Observatory (AUKR). The target spectra cover the wavelength range from 4000 to 7600~\AA.

The high resolution UVES spectra (R=~107,200) of LP~Ori was extracted from the ESO Science Portal \footnote{https://archive.eso.org/scienceportal/home}. These spectra cover the wavelengths regions of 3732 to 5000 ~\AA~ (UV to Blue) and 4583 to 6686 ~\AA ~(Visual to Red). The UVES spectrograph is attached on the 8.2 m VLT located at the Antofagasta Region. Furthermore, the ESPaDOnS Stokes V spectra, with resolving power of $\sim$65,000 and spectral
range of 3700–10400~\AA, were downloaded from the PolarBase archive\footnote{http://polarbase.irap.omp.eu/}. The ESPaDOnS 
spectropolarimeter is attached to the 3.6~m Canada–France–Hawaii Telescope (CFHT).

The data reduction of each spectra were performed through the pipelines developed especially for the observed instrument. Standard IRAF packages \citep{tody86,tody93} were used to calibrate the wavelength and normalize the continuum level of each scientific data. Furthermore, unblended and moderate lines on each scientific spectra were selected to calculate the heliocentric radial velocity and later this value was used to  shift the spectra of each target star. All of the spectral information are presented in Table~\ref{obsinfo}.

\begin{center}
\begin{table*}%
\footnotesize
\caption{Spectroscopic and observational information of LP~Ori.}
\centering
\begin{tabular}{llllllllll}
\hline \hline
$Spectrograph$	& $Whoppshel$	&	 $S/N$	&	 $V_{\rm r}$	&	$UVES $	&	 $S/N$	&	 $V_{\rm r}$	&	$EsPaDons$	&	 $S/N$	&	 $V_{\rm r}$ \\
\hline						
$R$	& $30,000$ & & $km\,s^{-1}$ &	$107,200$ &	& $km\,s^{-1}$ & $65,000$ & & $km\,s^{-1}$ \\
\hline							
$Star Name$	& & & &	$Obs. Date$ /$ Exposure Time (sec)$ & &	& & & \\
\hline
$LP$~Ori	& $ 2023.02.25$ / $3600$ & $120$ & $53\pm5.70$	& $2004.01.12$ / $580$	& $185/119$ & $3\pm0.50$ & $2006.01.12$  & $338$	& $21\pm2$ \\

& &   & 	& 	$2017.12.18$ / $1900$ & 110 & $32 \pm 1.00$ &  $2007.03.07$ &  $270$	& $21\pm2$  \\
 
	 &		&		&		& &  &  &	$2010.02.23$ 	&	$270$  &	$3\pm0.3$ 	\\

	 &		&		&		&  &&  &	$2012.12.29$ 	&	 $600$  &	 $3\pm0.3$	\\

	 &		&		&		&  & &  &	 $2013.11.26$	&	$ 640$ &	 $3\pm0.3$	\\
         
\hline
\bottomrule
\end{tabular}
\label{obsinfo}
\end{table*}
\end{center}

\begin{center}
\begin{table*}%
\caption{Atmospheric parameters and projected rotational velocity of LP~Ori.}
\centering
\begin{tabular}{l@{\hskip 0.11in}c@{\hskip 0.11in}c@{\hskip 0.11in}c@{\hskip 0.11in}c@{\hskip 0.11in}c@{\hskip 0.11in}c@{\hskip 0.11in}c@{\hskip 0.11in}c@{\hskip 0.11in}c@{\hskip 0.11in}c@{\hskip 0.11in}c@{\hskip 0.11in}c@{\hskip 0.11in}c@{\hskip 0.11in}c@{\hskip 0.11in}c@{\hskip 0.11in}c@{\hskip 0.11in}}
\hline \hline
\multicolumn{1}{c}{} & \multicolumn{2}{c}{Si} & \multicolumn{2}{c}{H$_{\gamma}$ } & \multicolumn{4}{c}{adopted}	\\
\hline
Star~Name &  $T_{\rm eff}$ & log\,\textsl{g} & $T_{\rm eff}$ & log\,\textsl{g} & $T_{\rm eff}$ & log\,\textsl{g} & $\xi$ & $v$\,sin\,$i$ \\
           & (K) & (cm\,s$^{-2}$) & K & (cm\,s$^{-2}$) & (K) & (cm\,s$^{-2}$) & (km\,s$^{-1}$) & (km\,s$^{-1}$)\\
\hline
LP~Ori & $22,000\pm1,000$ & $4.50\pm0.25$ & $21,000\pm1,000$ & $4.50\pm0.25$ & $22,000\pm1,000$ & $4.50\pm0.25$ & $6.7\pm1.4$ & $90\pm5$	\\
\bottomrule
\end{tabular}
\label{atmpar2}
\end{table*}
\end{center}

\section{Spectral Analysis}
\subsection{Atmospheric parameters and abundance analysis}
\label{atmos}

\begin{figure}
\begin{center}
\hbox{\hspace{0cm}\includegraphics[width=\columnwidth]{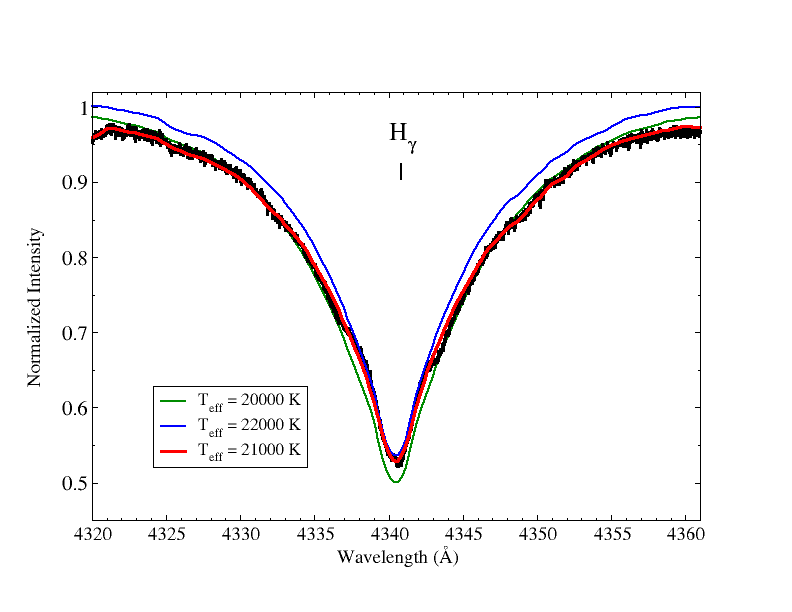}}
\caption{Comparison of the synthetic H$\gamma$ Balmer line profile fits to the UVES spectra of LP~Ori.}
\label{fig1}
\end{center}
\label{hgamma1}
\end{figure}

Non-LTE, plane-parallel, and hydrostatic model atmospheres for LP\,Ori was employed from \citet{lanz07}. We used the grids generated by TLUSTY \footnote{https://www.as.arizona.edu/hubeny/tlusty208-package/} and the synthetic spectra produced by  SYNSPEC54 code \citep{hubeny21}. The spectral lines of the synthetic spectra were broadened by the convolution of the rotational and microturbulent velocity, and the spectroscopic instrumental profile. The Kurucz’s atomic line list\footnote{http://kurucz.harvard.edu/linelists} was used 
as a source of atomic line parameters
for opacity calculations. Model atmospheres were generated for intervals of 1,000~K for $T_{\rm eff}$ changing from 16,000 to 30,000~K, and intervals of 0.25~dex changing from 3.75 to 4.75~dex for log~$g$. The initial atmospheric parameter of LP~Ori was derived by generating synthetic spectra of different pairs of $T_{\rm eff}$ and log~$g$, and adjusting these to the normalized observed H$\gamma$ Balmer line shown in Figure~\ref{fig1}.

\begin{figure}
\begin{center}
\hbox{\hspace{0cm}\includegraphics[width=\columnwidth]{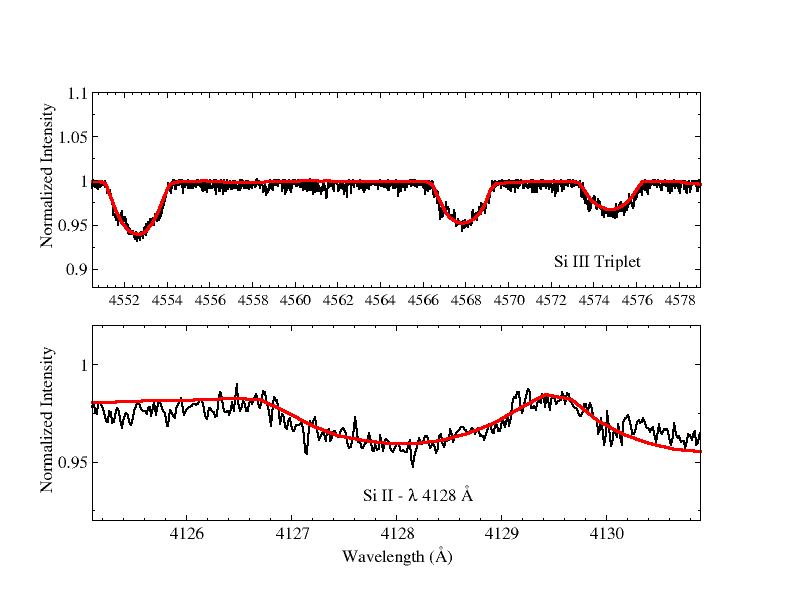}}
\caption{Si~II and Si~III multiplet lines of LP~Ori.}
\label{silines}
\end{center}
\label{hgamma2}
\end{figure}

Later, the ionization equilibrium of the two stages of silicon Si~II $\lambda$4128 / Si~III $\lambda$4552 for LP~Ori were compared to the observed values to refine the $T_{\rm eff}$ and log~$g$. The synthetic spectra fit to the observed Si~III multiplet near 4560~\AA~ is presented in Figure~\ref{silines}. 

\begin{table*}
\centering
\caption{Comparison of the elemental abundances of $LP$~Ori relative to the Sun \citep{asplund09}. } 
\begin{tabular}{lcccccccccccccccccc}\hline\hline
\hline			
Ions &log$\epsilon$ 	&	[X/H]	&	N	& [X/H]$_{mean}$ & log $\epsilon_{\odot}$ \\
\hline										
He\,I	       &	$11.09 \pm 0.13$	&	0.16	&	11	&     0.16        &   10.93 \\
C\,II       &	$8.48\pm0.09$	&	0.05	&	6	&     0.05        &   8.43	\\
N\,II	       &	$7.86\pm0.06$	&	0.03	&	8	&     0.03        &   7.83	\\
O\,II	       &    $8.74\pm0.14$	&	0.05	&	22	&     0.05        &   8.69	\\
Mg\,II	       &	$7.55\pm0.10$	&	$-0.05$	&	1	&     $-0.05$     &   7.60	\\
Al\,II	       &	$6.28\pm0.12$	&	$-0.17$	&	3	&     $-0.17$     &   6.45	\\
Si\,II	       &	$7.71\pm0.08$	&	0.20	&	2	&     0.08        &   7.51	\\
Si\,III	       &	$7.55\pm0.08$	&	0.04	&	5	&0.08 	          &   7.51	\\
S\,II	       &	$7.12\pm0.14$	&	0	&	4	&     0.00        &   7.12	\\
Fe\,III	       &	$7.53\pm0.05$	&	0.03	&	4	&     0.03        &   7.50	\\
\hline\hline
\end{tabular}
\label{abu}
\end{table*}

The calculation of the microturbulence velocity and projected rotational velocity was derived by fine tuning the best fit of the observed Si~III multiplet lines (4552, 4567, and 4574~\AA) for  LP~Ori in Figure~\ref{silines}. The atmospheric parameters measured from the hydrogen Balmer and silicon lines for LP~Ori are in good agreement, as seen in Table~\ref{atmpar2}.

\begin{figure}
\begin{center}
\includegraphics[scale=0.35]{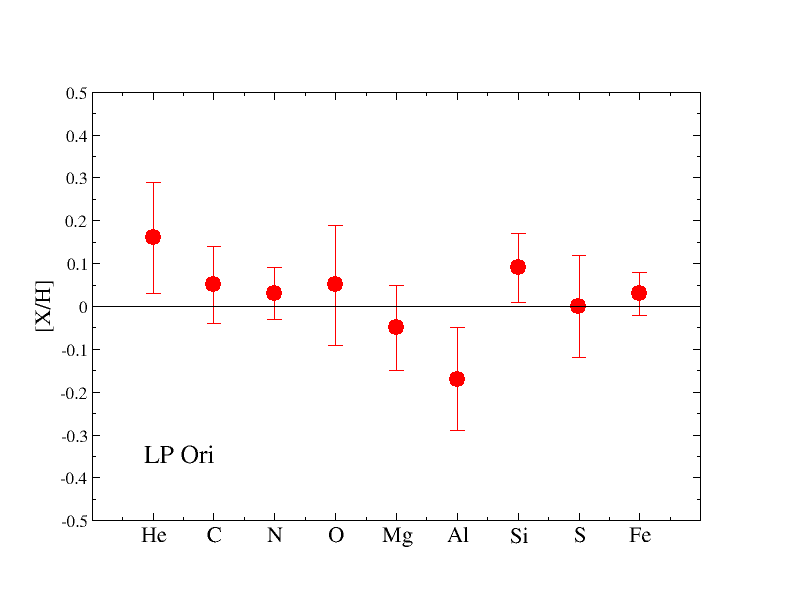}
\caption{Elemental abundance values relative to the Sun for LP~Ori. }
\label{abu1}
\end{center}
\end{figure}

Chemical abundance analysis was performed to the high-resolution UVES spectra of LP~Ori by adjusting the best fit of their synthetic spectra. The total uncertainties of each ion were calculated as described in \citet{elmasli23}. The nominal uncertainty value of 0.10 dex were given to the abundance measurements of single lines. The derived abundances of each ion within the total uncertainties of LP~Ori is listed in Table~\ref{abu}. We present in Figure~\ref{abu1} the elemental abundances of LP~Ori relative to the solar values adopted from \citet{asplund09}. In this figure, helium and silicon exhibit slight overabundance, while aluminum is slightly underabundant relative to the Sun. All the other elements (C, N, O, Mg, Si, S, and Fe) are close to solar abundances within the elemental error bars. 

He-rich CP4 stars are generally main-sequence and magnetic stars with helium abundance of $\sim$ 0.5 \citep{zboril97}. \citet{osmer74} note that the abundance of helium-rich B-type CP4 stars, $(He)/[\textit{N}(H)+\textit{N}(He)]$, ranges approximately from 0.2 to 0.9, whereas it is 0.1 for normal B-types. This helium abundance value of LP~Ori is calculated as 0.12, falling below the minimum threshold of 0.2. Even though LP~Ori has a magnetic field, the helium abundance is not as high as a CP4 star.  

The early B-type stars in the $\lambda$~Orionis Group \citep{elmasli23}, which is also part of the Orion star-forming complex, exhibits similar Al abundance value to that of LP~Ori. Therefore, a plausible interpretation is that the proginator stars exhibited a slight deficiency in their 
 aluminum abundances. 

\subsection{Balmer emission line profiles}

\begin{figure*}
\begin{center}
\hbox{\hspace{2.8cm}\includegraphics[scale=0.45]{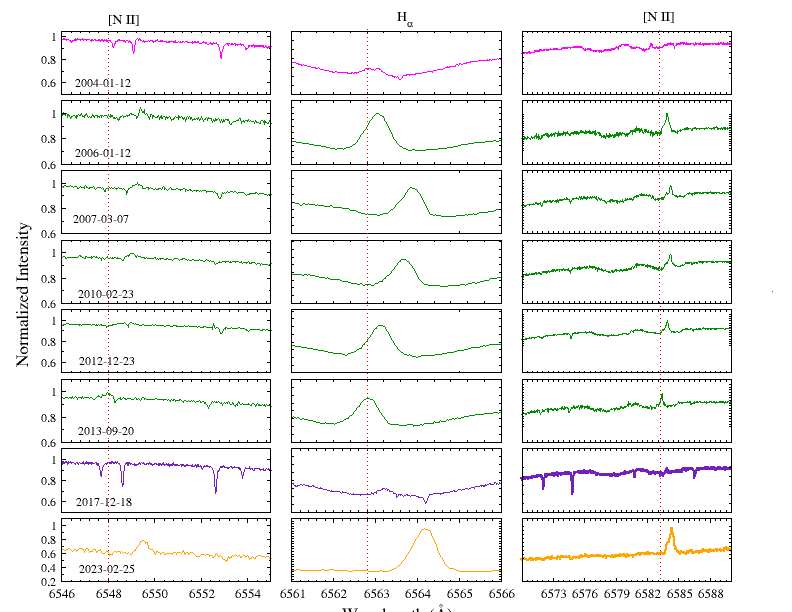}}
\caption{The H$\alpha$ line profiles and [NII] emission lines.}
\label{lporiha}
\end{center}
\end{figure*}

\begin{figure*}
\begin{center}
\hbox{\hspace{2.8cm}\includegraphics[scale=0.45]{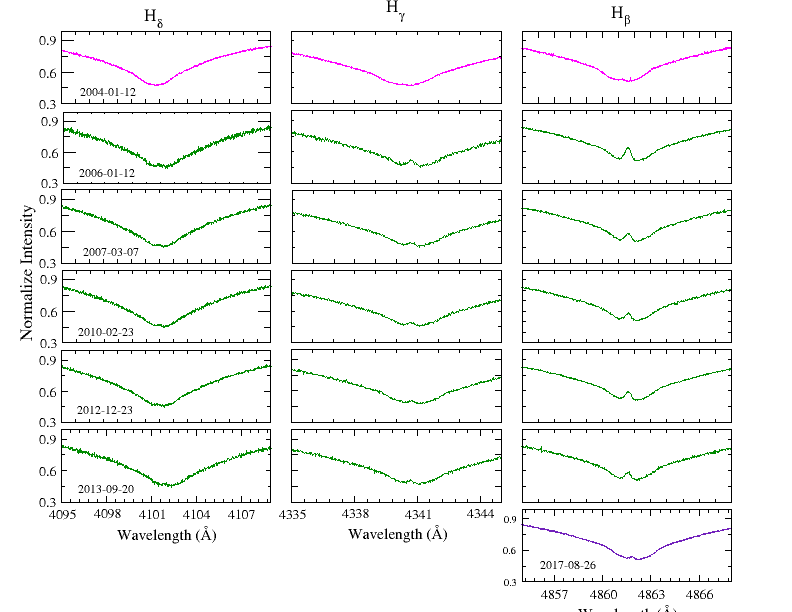}}
\caption{Spectral Balmer line profiles of LP~Ori.}
\label{lporibgd}
\end{center}
\end{figure*}

\begin{figure*}
\centering
\includegraphics[scale=0.45]{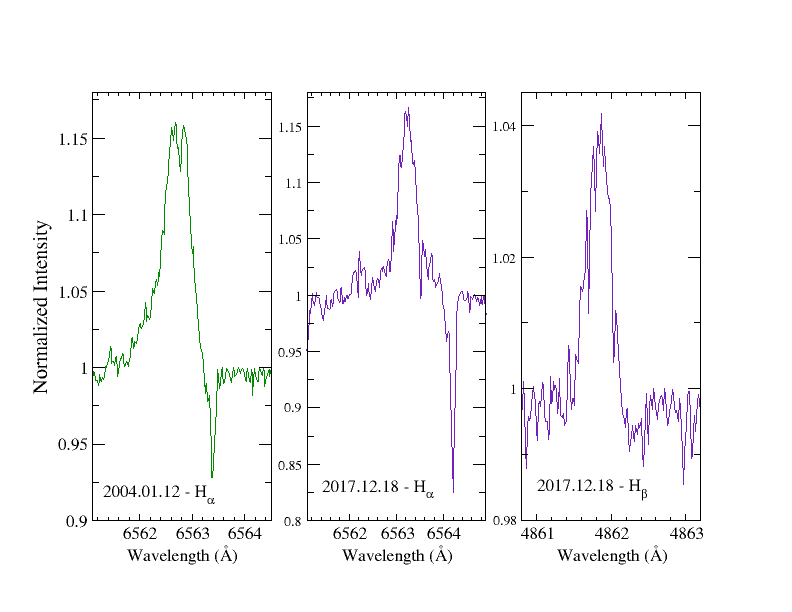}
\caption{Inverse P-Cygni in the core of the Balmer profiles.}
\label{pcygni}
\end{figure*}

All of the hydrogen Balmer line profiles seen in each spectra of LP~Ori are presented in Figure~\ref{lporiha} and ~\ref{lporibgd}. Notably, emission in the cores of the H$\alpha$ profiles are visible in all spectra, with the exception of those acquired in 2004 and 2017.
Moreover, all of the H$\beta$ and H$\gamma$ Balmer profiles in Figure~\ref{lporibgd} superimpose weaker emissions compared to the same dated H$\alpha$ profiles. Thus, the H$\alpha$ profiles of LP~Ori for the years 2006, 2007, 2010, 2012, 2013, and 2023 are classified as Type-I emission. In 2004 and 2017 inverse P-Cygni profile appear in the cores of the H$\alpha$ profiles as shown Figure~\ref{pcygni}. Thereby, the H$\alpha$  profiles for the years 2004 and 2017 are classified as Type-IV-R. 

\begin{figure*}
\begin{center}
\hbox{\hspace{3.2cm}\includegraphics[scale=0.45]{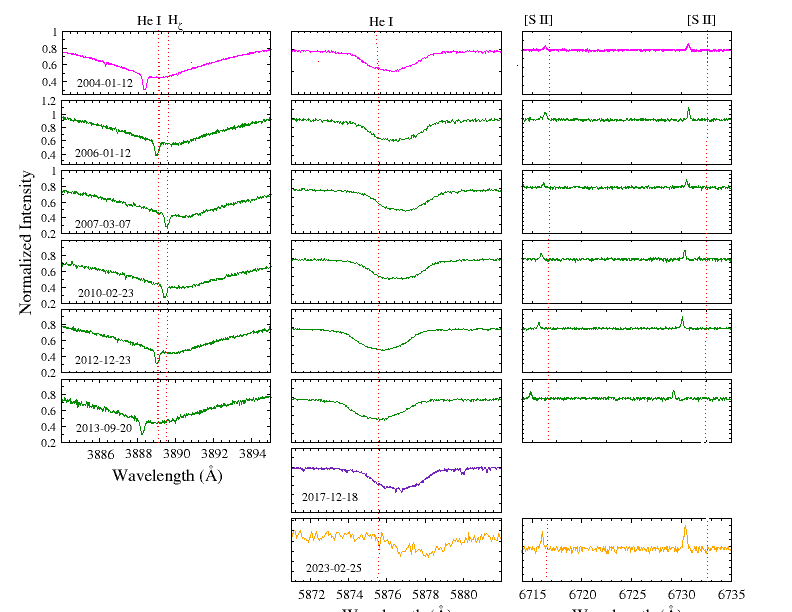}}
\caption{He\,I line profiles and [SII] emissions lines.}
\label{lporishe}
\end{center}
\end{figure*}

The forbidden lines of nitrogen at 6548/6583~\AA~ and sulfur at 6716/6731~\AA~are also visible in the Figures~\ref{lporiha} and~\ref{lporishe}, respectively. The He\,I absorption lines at $\lambda$3888 and $\lambda$5875 in each spectra are shown in Figure~\ref{lporishe}. In the year 2023 an emission is visible in the center of the He\,I $\lambda$5875 absorption line.
 
 \subsection{Non-photospheric spectral lines}

The non-photospheric Ca HK lines visible in the spectra of the Herbig Ae/Be stars may originate from its circumstellar disc or the interstellar medium. In order to clarify this we used The Local Interstellar Medium (LISM) \footnote{http://lism.wesleyan.edu/LISMdynamics.html} which calculates the radial and transverse velocity of the LISM and furthermore lists the Local Interstellar Clouds towards the line of sight of the target star. The velocity vectors of the local ISM ($V_{ISM}$) towards the line of sight of LP~Ori is calculated as 23.821 km\,s$^{-1}$.

If the radial velocity measurements of the Na\,D and Ca\,II absorption lines closely match the target values, then the origin is accepted as the circumstellar disc around the Herbig Ae/Be star, as described in \citet{rebollido20}. If the $V_{Ca\,HK}$ and/or $V_{Na\,ID}$ values are close to $V_{ISM}$ than the origin is accepted as ISM. If the time-series Ca\,II~HK and/or Na\,ID non-photospheric absorption lines show redshifts close to $\sim$20~km\,s$^{-1}$ from time to time and later blueshifts than there are exocomets in their CS enviroment.

\begin{table*}
\centering
\caption{ The radial velocity values of the interstellar medium lines.}
\begin{tabular}{l@{\hskip 0.10in}c@{\hskip 0.10in}c@{\hskip 0.10in}c@{\hskip 0.10in}c@{\hskip 0.10in}}
\hline 
\hline			
LP\,Ori & & & & \\
\hline								
$date$ & 2023  & $2004$ / $2017$	& $2006$ /$2007 $ / $2010$ / $2012$ / $2013$ &	\\
\hline								
~He\,I ($\lambda$~3888~\AA)   & …	   & $-55$ / ...  & $-25$ / $-38$ / $-37$ / $17$ / $11$ & \\
~Ca\,I $-$ K	           & …	   & $-16$ / ...  & 10 / $-3$ / $-2$ / 19 / 46	  & \\
~Ca\,II $-$ H                 & …	   & $-14$ / ...  & 12 / $-1$ / $-0.1$ / 20 / 48  & \\
~Na\,I ($\lambda$~5889~\AA)   & 32	   & ... / 17	  & 5 / $-8$ / $-7$ / 14 / 41	  & \\
~Na\,I ($\lambda$~5895~\AA)   & 32	   & ... / 17	  & 5 / $-8$ / $-7$ / 14 / 41	  & \\
~[N\,II] ($\lambda$~6548~\AA) & 35    & ...          & 6 / $-6$ / $-4$ / 16 / 43	  & \\
~[N\,II] ($\lambda$~6583~\AA) & 29    & ...          & 5 / $-9$ / $-6$ / 14 / 41     & \\
~[S\,II] ($\lambda$~6716~\AA) & 33	   & ... / -8     & 6 / $-6$ / $-8$ / 16 / 43	  & \\
~[S\,II] ($\lambda$~6731~\AA) & $-56$ & ... / $-68$  & $-81$ / $-94$ / $-93$ / $-73$ / $-44$ & \\
\hline \hline
\end{tabular}
\label{ism}

\end{table*}

\begin{figure*}
\centering
\hbox{\hspace{2.2cm}\includegraphics[scale=0.55]{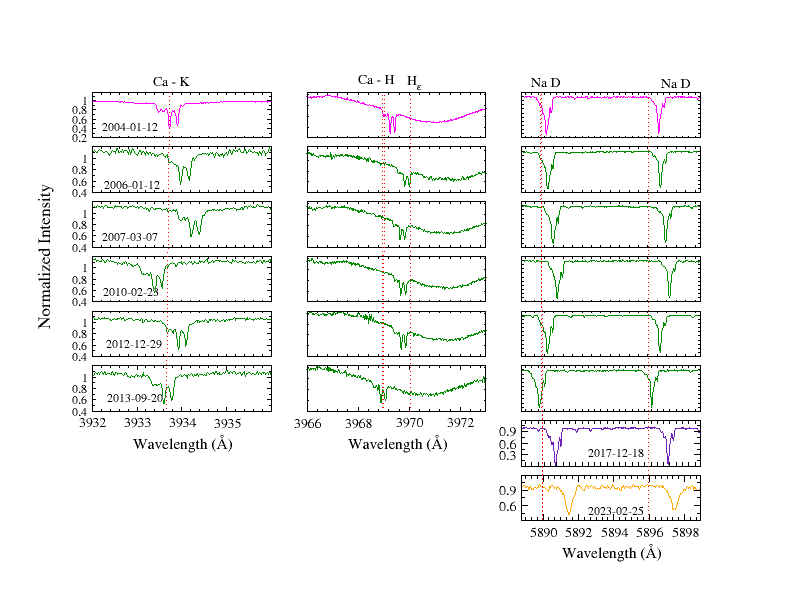}}
\caption{The Ca HK and Na D doublet lines of LP\,Ori.}
\label{lporicana}
\end{figure*}

We first compared the stellar $V_{\rm r}$ of each spectra presented in Table~\ref{obsinfo} with the $V_{Ca~HK}$ and $V_{Na~ID}$ values of LP~Ori in Table~\ref{ism}. This comparison inferred that the $V_{\rm r}$ of LP~Ori is different than CS enviroment. 

To decide if the CS origin is ISM; comparisons of the radial velocities show that in the spectra of LP~Ori both $V_{Ca~H}$ and $V_{Ca~K}$  are different than $V_{ISM}$.

The non-photospheric Ca~H\&K and Na~I~D lines in the irregular dated spectra of $LP$~Ori from 2004 to 2023 (as shown in Figure~\ref{lporicana}) show red- and blueshifts which may be caused by exocomets and/or planetesimals in the CS enviroment of the Herbig Ae/Be nature.

\begin{figure*}
\centering
\hbox{\hspace{2.2cm}\includegraphics[scale=0.55]{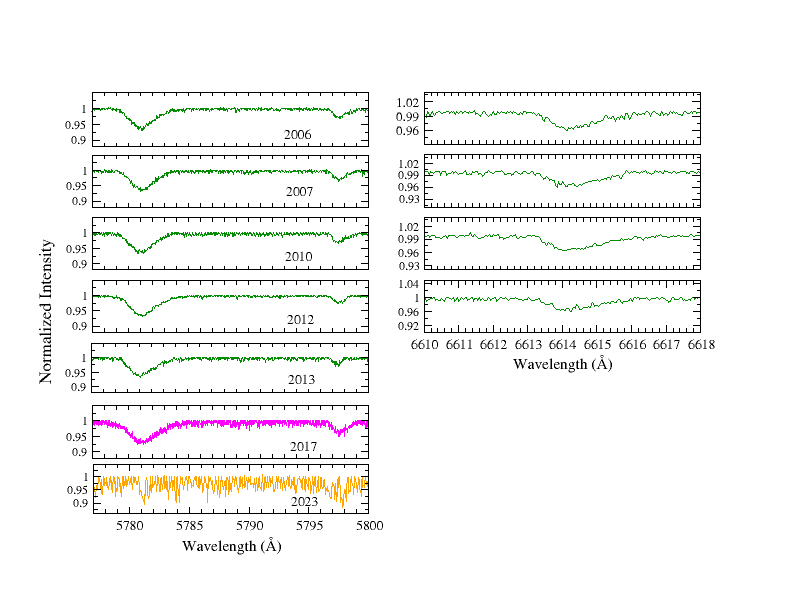}}
\caption{DIBs present at $\lambda$5780.5, 5797.1, and 6613.6~\AA.}
\label{diffuse}
\end{figure*}

Additional non-photospheric spectral features visible in the spectra of LP~Ori are the strong Diffuse Interstellar Bands (DIBs), presented in Figure~\ref{diffuse}, at $\lambda$5780.5, 5797.1, and 6613.6~\AA. The central
wavelengths, the widths (FWHM), and the equivalent widths (EW) of each band are listed in Table~\ref{dibs}. The origins of DIBs can be attributed to the presence of large molecules generated within the atmospheres of dying stars and/or Polycyclic Aromatic Hydrocarbons situated within highly dense interstellar regions associated with star-forming environments \citep{york24}.

\begin{table}
	\centering
	\caption{Diffuse Interstellar Bands in the Spectra of LP~Ori.}
	\label{abutab}
	\begin{tabular}{lccccl} 
		\hline\hline
		$\lambda$   & Date   & FWHM   & EW \\
		 (\AA)      & (Year) & (\AA)  & (m\AA) \\
		\hline\hline
	         5780.6   & 2006 & 2.542 & 14.41 \\ 
                          & 2007 & 2.372 & 16.04 \\
                          & 2010 & 2.382 & 16.08 \\
                          & 2012 & 2.573 & 18.72 \\
                          & 2013 & 2.406 & 15.08 \\
                          & 2017 & 2.815 & 20.14 \\
                          & 2023 & 0.487 &  4.74 \\
                  \hline
            
                 5797.1   & 2006 & 1.137 &  3.45 \\
                          & 2007 & 1.188 &  3.96 \\
                          & 2010 & 1.148 &  4.00 \\
                          & 2012 & 1.219 &  3.19 \\
                          & 2013 & 0.912 &  6.30 \\
                          & 2017 & 1.337 &  5.60 \\
                          & 2023 & 0.565 &  6.20 \\
                  \hline
                 6613.7   & 2007 & 1.670 &  6.15 \\
                          & 2010 & 1.878 &  7.00 \\
                          & 2012 & 1.776 &  6.30 \\
                          & 2013 & 1.855 &  6.70 \\
		\hline\hline
	\end{tabular}
 \label{dibs}
\end{table}

\section{Spectral Energy Distribution}
\subsection{Excess Infrared Radiation}

IR-excess refers to a measurement indicating that a star emits more IR radiation in its SED than would be expected if it behaved like a blackbody radiator. This excess IR radiation is often caused by the presence of circumstellar dust, which gets heated by the star's light and then re-emits radiation at longer wavelengths. Such occurrences are commonly observed in young stellar objects and stars that have evolved beyond the asymptotic giant branch stage.

Regarding the star's circumstellar characteristics, for many Herbig Ae/Be stars, accretion rates from the near-stellar disk to the star are calculated either directly by modelling the near-UV radiation excess \citep{2015MNRAS.453..976F} or indirectly by correlations with spectral emission lines \citep{2019AJ....157..159A}. However, the techniques used are significantly affected by observational biases. Therefore, it is necessary to rely mainly on the analysis of SEDs at IR and longer wavelengths \citep{1992ApJ...397..613H}. Studies of young stars' SED shape have proven very useful for determining their nature and evolutionary status \citep{1998ApJ...497..721N}. Since the spectrum emitted from a young star depends on the distribution and physical properties of the surrounding dust and gas, it is natural to expect the shape of the SED to depend on the evolutionary state of the source. A protostar embedded in the main cloud should have a different IR signature than that of a more evolved pre-main sequence star, in which most of the circumstellar matter is concentrated on the central star.

Therefore, SEDs obtained mainly in the near and mid-IR regions are used to understand the star formation process. IR SEDs are classified according to the comparison of radiation from a dust disk with that of a star. The classification scheme, based on the slope of the SED at 2.2 $\mu m$, is the most successful among such classifications \citep{1989ApJ...340..823W}. The SED is characterized by a spectral index given by $\alpha = -\mathrm{d}\log(\lambda F_{\lambda}) / \mathrm{d}\log (\lambda$) where $F_{\lambda}$ denotes the flux density at wavelength $\lambda$. The spectral index $\alpha$ is usually measured in the wavelength range of $\alpha$ = 2.2 $\mu m$ and 10-25 $\mu m$. \citet{1989ApJ...340..823W} use the slope between 2.2 $\mu m$ and the longest observed wavelength between 10 and 25 $\mu m$ to calculate the index. \citet{1987IAUS..115....1L} explains the evolutionary condition of such an object by referring to its SED and categorizes it into type I, II, and III, as illustrated in \citet{2011PhDT.......323C}. Type I is characterized by a compact accretion disc and a large infalling envelope with a bipolar cavity. Emission from the envelope dominates its SED, resulting in a broader distribution than typical blackbody curve, with an increase longward of 2 $\mu m$. The envelope is about 0.1 M$_{\odot}$ at this stage. The age of the object is about $10^5$ years. Type II object has largely dissipated its envelope and is surrounded by a flared disc. The dominant components in the SED are emissions from the central source and the disc, leading to a broader distribution compared to blackbody spectra. Beyond 2 $\mu m$, the SED shows either a flat or negative trend. The disk mass is roughly 0.01 M$_{\odot}$ and the age of the object is about $10^6$ years. Type III object has mostly cleared its disc, so its SED reflects a reddened stellar photosphere of a star that is either very close to or on the Zero Age Main Sequence (ZAMS). In the Near-IR, there is minimal or no excess radiation. The disk mass is roughly 0.003 M$_{\odot}$. The age of the object is about $10^7$ years \citep{2009ApJ...695..511C}.

Therefore, by examining the SEDs of the stars, it can be directly determined whether there are dust and disk structures around, indicating that they are still young objects, such as Herbig Ae/Be stars.

\subsection{Extinction and Reddening}
\label{extinction}

Dust plays a critical role within galaxies as it alters the observable characteristics by both absorbing and scattering starlight. It also emits the absorbed energy in the mid- and far-IR regions. Dust extinction, or the reduction of light intensity due to dust, generally decreases as the wavelength increases, meaning that shorter wavelengths experience more significant extinction. As a consequence, SEDs of galaxies take on a redder appearance, commonly referred to as 'reddening'. Therefore, it is essential to consider the effects of dust when interpreting the observations of different regions of the galaxy \citep{2012MNRAS.421..486X}. 

The extinction is predominantly seen in dark nebulae in our galaxy and known as the loss of light at different wavelengths. In this sense, $A_{\lambda}$ is the total extinction at the wavelength ${\lambda}$ and selective extinction, on the other hand, refers to the difference between the extinctions at different wavelengths and given as;
\begin{equation}
    A_{\mathrm{B}}-A_{\mathrm{V}}=(\mathrm{B}-\mathrm{V})-(\mathrm{B}-\mathrm{V})_0=\mathrm{E}(\mathrm{B}-\mathrm{V})
	\label{eq:selective_ext}
\end{equation}

where $(\mathrm{B}-\mathrm{V})$ is apparent colour index, $(\mathrm{B}-\mathrm{V})_0$ is intrinsic colour index and $\mathrm{E}(\mathrm{B}-\mathrm{V})$ is colour excess. In order to deredden a spectrum or a SED, the extinction profile corresponding to a related $\mathrm{E}(\mathrm{B}-\mathrm{V})$ value must be known. The generalised colour excess, which describes the extinction variation as a function of wavelength and known as dereddening profile, is;
\begin{equation}
   \mathrm{E}(\mathrm{\lambda}-\mathrm{V})=A_{\mathrm{\lambda}}-A_{\mathrm{V}}
	\label{eq:gen_col_exc}
\end{equation}

Besides, the ratio of total extinction to selective extinction is given as;
\begin{equation}
   R_{\mathrm{V}}=A_{\mathrm{V}}/\mathrm{E}(\mathrm{B}-\mathrm{V})
	\label{eq:tot-to-sel-rat}
\end{equation}

The $R_{\mathrm{V}}$ parameter here characterises the dust properties in the region that produces the extinction ranges from about 2.0 to about 5.5 with a typical value of 3.1 from diffuse to dense interstellar medium \citep{2004AcA....54..375G}. However, in dense molecular clouds $R_{\mathrm{V}}$ can change from 5 to 7 \citep{1989ApJ...345..245C}. 

The extinction law describes the change of relative extinction $A_{\mathrm{\lambda}}/A_{\mathrm{V}}$ with wavelength $\lambda$ and provides an opportunity to obtain the extinction in some spectral region depending on the extinction in a different spectral region \citep{2004AcA....54..375G}. It is given as;
\begin{equation}
 A_{\mathrm{\lambda}}/A_{\mathrm{V}}=a(x)+b(x).R_{\mathrm{V}}^{-1}
	\label{ext_law}
\end{equation}
where $x(\mu m^{-1}) = 1/{\lambda}$, and $a(x)$ and $b(x)$ are the wavelength dependent coefficients. In the Equ.~\ref{ext_law}, the coefficients are calculated as given in \citet{1989ApJ...345..245C};\\

\hspace{-1.2em} For IR: $0.3~\mu m^{-1} \leq x \leq 1.1~\mu m^{-1}$
\begin{equation*}
\begin{aligned}
&a(x)= \;\,\, 0.574x^{1.61}\\
&b(x)=-0.527x^{1.61}
	\label{IR}
\end{aligned}
\end{equation*}\\

\hspace{-1.2em} For Optical/Near IR: $1.1~\mu m^{-1} \leq x \leq 3.3~\mu m^{-1}$ and $y=(x-1.82)$
\begin{equation*}
\begin{aligned}
\;\; &a(x)=1+0.17699y-0.50447y^2-0.02427y^3+0.72085y^4\\
&\qquad \quad +0.01979y^5-0.77530y^6+0.32999y^7\\
\;\; &b(x)=1.41338y+2.28305y^2+1.07233y^3-5.38434y^4\\
&\qquad \quad -0.62251y^5+5.30260y^6-2.09002y^7
	\label{OP-NIR}
\end{aligned}
\end{equation*}

Even though different polynomials to derive $a(x)$ and $b(x)$ coefficients are given for the UV region between $3.3~\mu m^{-1} \leq x \leq 8.0~\mu m^{-1}$ in \citet{1989ApJ...345..245C}, these polynomials are not mentioned here since there is no SED value in the UV region in this study.

If $\mathrm{E}(\mathrm{B}-\mathrm{V})$ value of the target star is known and $R_{\mathrm{V}}$ can be assumed, the total extinction $A_{\mathrm{V}}$ value in the V band can be found from Eqn.~\ref{eq:tot-to-sel-rat}. Once $A_{\mathrm{V}}$ is known, $A_{\lambda}$ can be calculated from the extinction law. Thus, the dereddening profile can be obtained from Eqn.~\ref{eq:gen_col_exc}. As a result, the total extinction in the spectral flux density can be corrected using $A_{\mathrm{V}}$ and the reddening effect can be corrected through $\mathrm{E}(\mathrm{\lambda}-\mathrm{V})$. Basically, these corrections are made by multiplying the spectrum by $10^{\left[ 0.4 \times A_{\mathrm{V}}\right]}$ and $10^{\left[0.4 \times \mathrm{E}(\mathrm{\lambda}-\mathrm{V})\right]}$.

\subsection{SED Analyses of LP~Ori}

In the light of this information, SED data of  LP\,Ori were obtained and examined whether it shows excess radiation in the infrared region. If any evidence of IR-excesses is found, it can be taken as an indication that the stars are young and have circumstellar disks. 

\begin{figure}
\hbox{\hspace{0cm}\includegraphics[width=\columnwidth]{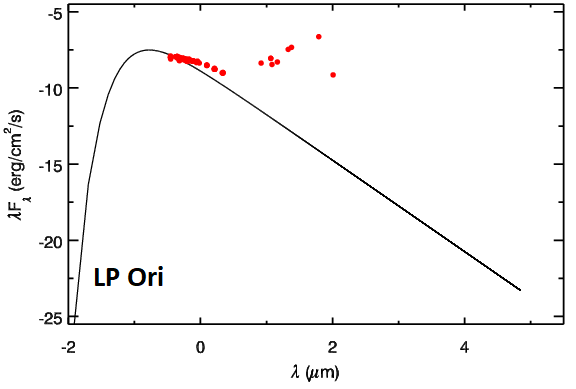}}
    \caption{Log-log plot of SED data for LP\,Ori. The red dots show the SED data, and the black solid line shows the Planck curves produced according to the temperature of the star given in Table~\ref{hrtable}. Since the errors of the data are too small, they are not included in the plots.}
    \label{fig:herbig-raw-planck}
\end{figure}

The SED data of LP\,Ori  was retrieved from CDS Portal\footnote{http://cdsportal.u-strasbg.fr/} and transferred to the log-log graphics which is given in Figure~\ref{fig:herbig-raw-planck}. In the graphic, an attempt was made to fit the SED data by applying the Planck curve generated based on the effective temperature.

In the graphs, the x-axis is the wavelength ($\lambda$) in $\mu m$, while the y-axis is the flux density ($\lambda \mathrm{F}_{\lambda}$) in $\mathrm{erg~cm^{-2}~s^{-1}}$ units. Although the excess radiations in the IR region is evident for LP\,Ori, it is seen in the graph that the data in the range of $-1 < \log{\lambda} < 1$, which corresponds to the visual region, does not match the blackbody curve. Considering that the star is in a star forming region, it is obvious that there will be a significant amount of gas and dust in the line of sight between the star and the observer. Therefore, it is understood that the data in the visual region is subject to a significant extinction.

The extinction and reddening effects have been eliminated in the light of the information given in Section~\ref{extinction}. The extinction and reddening values for LP\,Ori are given in Table~\ref{hrtable}.

\begin{figure}
\hbox{\hspace{0cm}\includegraphics[width=\columnwidth]{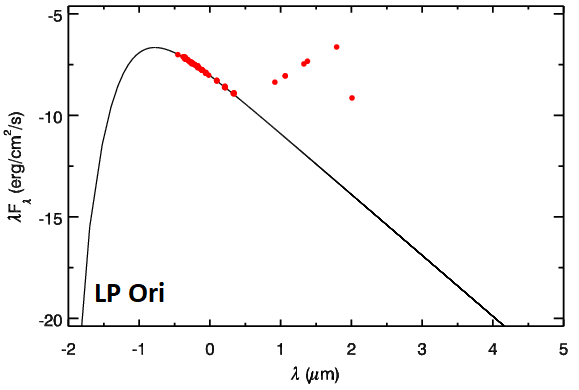}}
    \caption{Corrected SED data of LP~Ori.}
    \label{fig:herbig-corr-planck}
\end{figure}

Wavelength - Flux density graph obtained after the extinction and reddening corrections is shown in Figure~\ref{fig:herbig-corr-planck}. It should be noted that the corrected data are quite compatible with the Planck curves.

\begin{center}
\begin{table*}%
\caption{The parameters used to plot the position of LP~Ori on the H-R diagram.\label{hrtable}}
\centering
\begin{tabular}{lccccccccccc}
\hline\hline
 $Stars Name$ &	$m_{\rm{v}}$ &	$\pi$ &	$M_{\rm v}$	& E(B-V)	& R$_{\rm{v}}$	& BC & log($T_{\rm{eff}})$	& log($L$/L$_{\rm{\odot}})$ &		\\
 &	(mag)	&	(mas)	&	(mag)	&	(mag)	&	(mag)	&	(mag)	&	(K)	&	(cgs)	&			& \\
\hline												
$LP$~Ori & $8.46\pm0.009^a$ & $2.45\pm 0.02^b$ & $-1.99\pm0.01$ & $0.32^d$ & 5.25 & $-2.189^c$ & $4.34\pm0.02$ &	$3.24\pm0.08$ &	 \\
\hline\hline
\end{tabular}
\begin{tablenotes}
\item \citet{2002yCat.2237....0D}~$^a$, \citet{2020yCat.1350....0G}~$^b$, \citet{2013A&A...550A..26N}~$^c$, \citet{2013MNRAS.429.1001A}~$^d$,  
\citet{2016AcA....66..469K}~$^e$
\end{tablenotes}
\end{table*}
\end{center}

This graph shows that the target star shows excess radiation in its region. According to the SED shape, for LP\,Ori, the curve shows a substantial rise at around $\lambda = 3.16~ \mu m$ ($\log{\lambda} = 0.5 \mu m^{-1}$). This is an indicator of a compact accretion disk and an infalling envelope. Indeed it is stated that the star is located in the center of a CO shell \citep{2013AJ....146...85H}. So LP~Ori is still in the same environment as the cloud it was born in and its light is blocked by this CO shell. Therefore, it can be concluded that LP\,Ori is young and a Type I object having the age of around $10^5$ years.

\section{Evolutionary Status}
\label{evo}
\begin{figure}
\begin{center}
\hbox{\hspace{0cm}\includegraphics[width=\columnwidth]{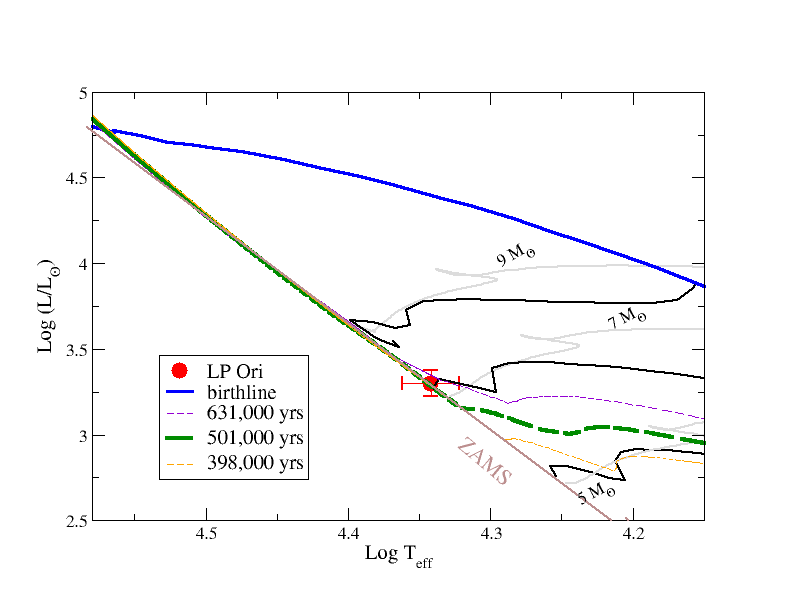}}
\caption{Location of LP~Ori on the H-R diagram.}
\label{hr2}
\end{center}
\end{figure}

LP\,Ori is plotted on the H-R diagram in Figure~\ref{hr2}. The spectroscopically derived $T_{\rm eff}$ values of LP\,Ori is withdrawn from Section~\ref{atmos}. Since the target is in the Orion Nebula, the extinction factor was calculated by using the formulas presented in Section~\ref{extinction}. The $R_{\rm V}$ and the color excess values used to calculate the reddening in the V band is presented in Table~\ref{hrtable}.
The V magnitude, measured by \citet{2002yCat.2237....0D}, the stellar distance  obtained from GAIA \citet{2020yCat.1350....0G}, the bolometric correction value from \citet{2013A&A...550A..26N}, and the extinction along the
line-of-sight were used to calculate the bolometric absolute magnitude of LP~Ori.

LP\,Ori is located between the zero-age main-sequence (ZAMS, brown line) and the birthline (blue) \citet{bressan12} as seen in Figure~\ref{hr2}, which is an indication that LP\,Ori is as Herbig Ae/Be star.
We plotted the PMS (solid black, solid grey evolution after ZAMS) evolutionary tracks from \citet{haemmer19} to determine the mass of LP\,Ori, considering its position on the H-R diagram. The location of LP\,Ori is just on the pre-mainsequence evolutionary track of 7$M_{\odot}$ which coincides with ZAMS. Thus, the mass of  LP\,Ori from Figure~\ref{hr2} is estimated as $7.0\pm1.7~M_{\odot}$. The isochrones from \citet{haemmer19} were also drawn on the H-R diagram to estimate the age of LP\,Ori as $501,000 \pm70,000$ years.

\section{Photometric Analysis}

 In this context, the photometric data of the target star was examined in order to understand whether they show any light variation and to investigate the cause of these changes. Accordingly, the photometric data was obtained from the TESS archives. Even though TESS has observed more than 64 sectors in the sky thus far, the data for LP\,Ori was only derived from sector \#32.

  \begin{figure}
\hbox{\hspace{0cm}\includegraphics[width=\columnwidth]{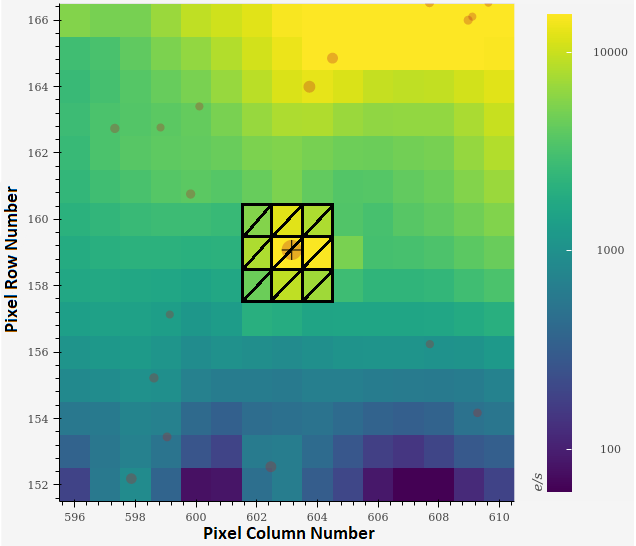}}
    \caption{All the reliable source locations plotted over the TESS Sector \#32 Target Pixel File and the mask used to extract the light curve.}
    \label{fig:gaia}
\end{figure}

Given TESS's wide field of view and the positining of LP\,Ori within a star-forming region, care was taken to ascertain the absence of contamination during the acquisition of the star's light curve. For this reason, a specific aperture mask, generated by the SPOC pipeline, was used to extract the light curve. Figure~\ref{fig:gaia} displays all the reliable source locations plotted over the TESS Sector \#32 Target Pixel File and the mask is shown with black hatched regions. As can be seen from the figure, there is no source in the mask that could cause contamination.

\begin{figure*}
\centering
\includegraphics[width=0.8\textwidth]{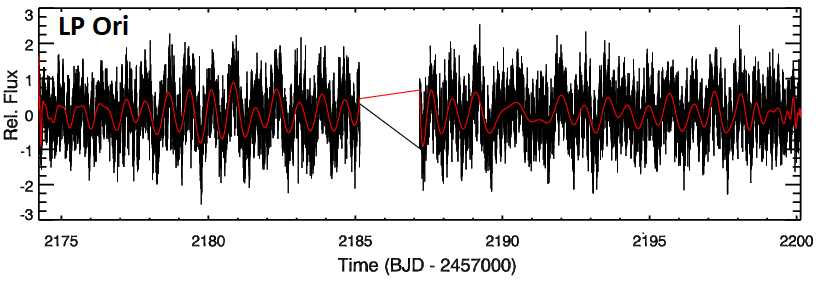}
    \caption{Light curves of LP~Ori from sector \#32. The red continues lines show smoothed light curves.}
    \label{fig:herbig-LC}
\end{figure*}

For the analyses, pre-search data conditioning simple aperture photometry (PDCSAP) flux was used in the study. Before the analyses, long term trends, small scale systematic trends, residuals from reaction wheel desaturation events and systematics from scattered light near the data downlink gap \citep{2021ApJS..257...53L} were corrected. The data covered observation intervals of approximately 27 days and consisted of photometric points taken every 120 seconds. Therefore, it enabled the detection of signals between a frequency resolution of about 0.036 cycles per day (c~d$^{-1}$) and Nyquist frequency of about 360 (c~d$^{-1}$). The light curve is presented in the Figure~\ref{fig:herbig-LC}. In the figure, the x-axis is the Barycentric Julian Day, and the y-axis is the normalised flux. In order to catch the changes more clearly, smoothed light curves shown with red continuous lines were also plotted on the data.

The time series were analysed by using the Lomb-Scargle algorithm \citep{1976Ap&SS..39..447L, 1982ApJ...263..835S} in IDL programming language\footnote{https://www.l3harrisgeospatial.com/Software-Technology/IDL}. An iterative process was applied to the light curves to detect the frequency and the amplitude of the highest peak.  For each amplitude spectrum, a local noise value was calculated for every 0.25 c~d$^{-1}$ frequency interval up to the Nyquist frequency, so that a variable noise profile was determined. Based on this noise characteristic, frequencies with the signal-to-noise ratio (SNR) > 5.037 were assumed to be significant \citep{2021AcA....71..113B}). 

Finally, the light curves were phased with the fundamental frequency obtained from the LS method and the brightness change was plotted against these phases. In this way, it was also investigated whether the light curves of the target stars could be represented by a sinusoidal structure originating from rotation.

\begin{figure}
\hbox{\hspace{0cm}\includegraphics[width=\columnwidth]{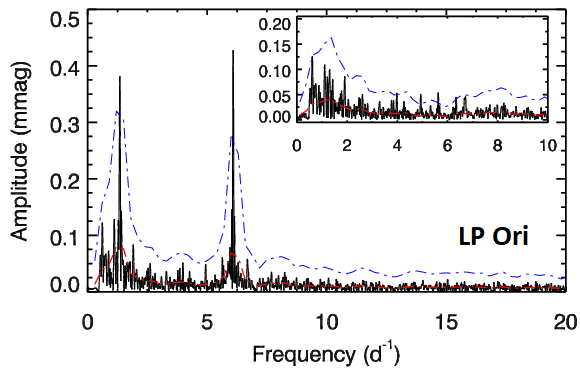}}
    \caption{Peridogram of LP~Ori. Large graphics show the low frequency region while small graphics display  high frequency regions. The red lines are SNR profiles of the periodograms for each 0.25 d$^{-1}$ frequency interval, and the blue lines are the significance levels at SNR = 5.037.}
    \label{fig:herbig-LS}
\end{figure}

\subsection{Frequency Analysis of LP~Ori}

In order to obtain information about these changes, frequency analyses were carried out using the Lomb-Scargle method. The amplitude spectra of the stars are given in Figure~\ref{fig:herbig-LS}.

\begin{figure}
\hbox{\vspace{0cm}\includegraphics[width=\columnwidth, height=0.225\textheight]{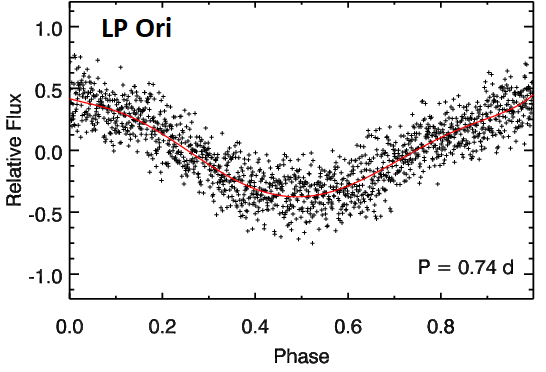}}
    \caption{Phased light curves of LP~Ori}
    \label{fig:herbig-phased}
\end{figure}

\begin{table}
	\centering
	\caption{Frequency, amplitude and SNR values of LP~Ori.}
	\label{tab:freq-table}
	\begin{tabular}{lccr} 
		\hline
		&Frequency& Amplitude & SNR  \\
        &d$^{-1}$& mmag &   \\
		\hline\hline
$f_1$&	6.085&	0.432&	13.29\\
$f_2$&	1.344&	0.380&	12.70\\
$f_3$&	0.614&	0.126&	4.98\\
$f_4$&	5.631&	0.052&	6.10\\
$f_5$&	3.983&	0.051&	4.14\\
$f_6$&	4.935&	0.051&	5.19\\
$f_7$&	6.713&	0.047&	4.16\\
$f_8$&	6.343&	0.045&	4.44\\

		\hline
	\end{tabular}
\end{table}

The periodogram of LP~Ori is frequency-rich (Table~\ref{tab:freq-table}). At first glance, two frequencies, $f_1$ = 0.164 days (6.085 d$^{-1}$) and $f_2$ = 0.744 days (1.344 d$^{-1}$), appear to be dominant in the periodogram (Figure~\ref{fig:herbig-LS}). As seen from the phased light curve in the Figure~\ref{fig:herbig-phased}, $f_2$ = 1.344 d$^{-1}$ of these frequencies produces a sinusoidal light curve. Therefore, the $f_2$ frequency seems to be related to the rotation of the star.

The observed change is attempted to be explained by the presence of a magnetosphere in certain stars with accretion disks, such as T Tauri and Herbig Ae/Be stars \citep{2021Univ....7..489P}. The magnetosphere is a region surrounding the star where the magnetic field controls the accretion of material from the disk onto the star. When a star possesses a strong magnetic field, the material transfer from the disk to the star is halted at a specific distance known as the Alfv{é}n radius (R$_A$). The region within this radius is called the magnetosphere, where the motion of the accreting gas is governed solely by the magnetic field. In this accretion scenario, some of the matter flows from the disk to the star along closed magnetic field lines near the magnetic poles, while the remaining matter escapes along open field lines at lower magnetic latitudes. Since the magnetic axis may not align with the rotational axis of the star, the geometry within the magnetosphere is not axisymmetric, leading to a spherical azimuthal inhomogeneity in the circumstellar environment around the star. Consequently, the star's rotation induces cyclic variability in the system, with a periodicity of either the rotational period ($P_{\mathrm{rot}}$) or half of it, depending on the visibility of one or both magnetic poles during a rotation period (P) \citep{2021Univ....7..489P}. Two main types of inhomogeneity contribute to generating this variability: (1) an asymmetric accretion stream inside the magnetosphere relative to the rotation axis and (2) a hot spot on the stellar surface where the accreted material impacts. Spectroscopy of lines originating near the disk/star interaction region can reveal the first item, while the second one can be identified through precise photometry and high-resolution spectroscopy, particularly using atmospheric lines sensitive to temperature \citep{2021Univ....7..489P}. Recently, a similar mechanism is also proposed for classical Be stars in \citet{2021ApJ...921....5B}, where 441 stars are analysed and it is revealed that such a mechanism shapes the periodogram of the star.

Additionally, when $f_1$ and $f_2$ frequencies are removed from the periodogram of LP~Ori, it is seen that there are five more frequencies (Table~\ref{tab:freq-table}) in the frequency range between 3 d$^{-1}$ and 7 d$^{-1}$ in the related graph in the Figure~\ref{fig:herbig-LS} . These changes indicate that the star may have a different nature, such that frequency interval typically falls within the range given for $\beta$~Cephei stars. This finding provides important evidence that the newborn star LP\,Ori may continue to live as a $\beta$~Cephei star.

\section{Discussion and Conclusions}

This is a thorough investigation of LP\,Ori located in the Orion Nebula. An up-to-date optical region spectra
 of LP\,Ori was obtained with T80 at AUKR while all previously acquired spectra, dated prior to the current observation, were retrieved from the archives of ESO and ESPaDOnS.

 Since LP\,Ori is located in M\,42 star-forming region, the origin of the emission profiles visible in all spectra need clarification, whether it is caused by its Herbig Ae/Be nature or influenced by the H\,II region. 
 Depending on the structure of the circumstellar disc of a Herbig Ae/Be stars, both symmetric and P-Cygni (and also inverse P-Cygni) emissions in the core of H$\alpha$ line may occur. The H\,II regions characteristic for H$\alpha$ emission is generally a narrow single-peaked symmetric line. Even though the H$\alpha$ emission EW of Herbig Ae/Be stars are generally larger (~4\AA) than the ones produced by H\,II regions (~1-2\AA), Hen 3-1121S and Hen 3-1121N \citep{carmona10} are two Herbig Ae/Be stars that exhibit weak emission components. Thus, the narrow emission ($\sim$1.4\AA) of LP\,Ori may be caused by the stars Herbig Ae/Be nature. While P-Cygni profiles visible in massive stars are caused by expanding envelopes, inverse P-Cygni profiles are due to outflowing envelopes. Thus, the inverse P-Cygni seen in the core of H$\alpha$ lines of LP\,Ori is the material outflowing from the source of the Herbig Ae/Be star towards its CS, rather than a H\,II region. One of the other obvious CS signature of Herbig Ae/Be stars is the emission in the $\lambda$5875 He\,I line, which is observable in the 2023 spectra (Figure~\ref{lporishe}). The absorption line of He\,I at $\lambda$3888 is superimposed on the H$\zeta$ line profile in each spectra in Figure~\ref{lporishe} and also the forbidden lines, can be attributed as outflows from the source Herbig Ae/Be star or the influence H~II regions on LP\,Ori. 

The majority of elemental abundances of LP\,Ori exhibit values proximate to solar levels within their uncertainties, with notable exceptions in the cases of helium and aluminum. 
 The magnetic characteristics and marginal helium overabundance observed initiate an examination into the classification of helium-rich CP4 star. Since the helium abundance value falls below the lower limit value indicative of helium-rich classification, it is more probable that LP\,Ori is a Herbig Ae/Be star.

 Furthermore, the TESS lightcurves of LP\,Ori indicated a rich frequency spectrum. The frequency analysis inferred LP~Ori as a newborn $\beta$~Cephei star. 

 CS enviroment around young Herbig Ae/Be stars as well as H~II regions are possible causes of IR-excess in the SED's of stars. 
 When LP\,Ori is considered as a Herbig Ae/Be star the SED's IR-excess character yield as a Type~I objects. Furthermore, the approximation of the age made with the SED for LP\,Ori ($\sim10^5$ year), is in good agreement with the one estimated from the isochrones in Section~\ref{evo}. Another piece of evidence referring to this target as a Herbig Ae/Be object is its position on the H-R diagram, located between the birthline and the ZAMS, as seen in the Figure~\ref{hr2}. 

Upon comprehensive consideration of these analyses collectively, it becomes more probable that LP\,Ori can be classified as a Herbig Ae/Be star, with its subsequent evolutionary phase suggesting a transition towards a ZAMS $\beta$~Cephei star. Further spectroscopic and photometric observations of LP\,Ori is required to understand the early stages of stellar evolution, particularly the processes involved in the formation and accretion of material around young and massive stars.

\bibliography{elmasli}

\section*{Acknowledgments}

The authors acknowledge the support of the Scientific and Technological Research Council of Turkey (T\"UB\.ITAK) through project 3501-121F426.

This article is based on observations made with the T80 - Prof. Dr. Berahitdin Albayrak Telescope, Ankara University Kreiken Observatory, Ankara, Turkey  with the project numbers 23A.T80.02. We extend special thanks to Jannat Mushreq Kam{\i}l AL-AZZAWI and Elif {\c{S}}ura ET{\.I}{\c{S}}KEN for their observation of the target star at the AUKR.

Based on data obtained from the ESO Science Archive Facility with DOI(s): https://doi.org/10.18727/archive/50.

Based on observations obtained at the Canada-France-Hawaii Telescope (CFHT) which is operated by the National Research Council of Canada, the Institut National des Sciences de lUnivers of the Centre National de la Recherche Scientique of France, and the University of Hawaii.

We extend our sincere appreciation to the anonymous referee for their thorough and insightful review of our manuscript.

\end{document}